# Controllable thermal radiation from twisted bilayer graphene

Yong-Mei Zhang[1], Mauro Antezza[2,3], and Jian-Sheng Wang[4]

[1]*College of Science, Nanjing University of Aeronautics and Astronautics, Nanjing 211106, People's Republic of China*

[2]*Laboratoire Charles Coulomb (L2C), UMR 5221 CNRS-Université de Montpellier, F-34095 Montpellier, France*

[3]*Institut Universitaire de France, 1 rue Descartes, F-75231 Paris, France*

[4]*Department of Physics, National University of Singapore, Singapore 117551, Republic of Singapore*

The presence of interlayer interactions in twisted bilayer graphene (TBG) enhances several characteristics, including the optical and electrical properties. We theoretically investigate the magic angle of TBG according to the vanishing of Fermi velocity and find double magic angles in a series. The thermal radiation from TBG can be tuned to the far infrared range by changing twist angles. The peculiar radiation spectrum is out of atmospheric window, which can be of great use in invisibility and keeping warm. The total radiation of TBG is slightly more than twice of a single layer graphene.

## I. Introduction

The twisted bilayer graphene (TBG) superlattice is formed by stacking two layers of graphene and twisting a small angle. The interlayer interactions in TBG generate extraordinary optical and electronic properties in such system, which attracts great interest from researchers both theoretically and experimentally[1-3]. The studies on TBG have been exciting and challenging thus far, especially after superconductivity was reported at the magic angle[4, 5].

Due to the unique structure of bilayer graphene, the twist angle is regarded as a controlled degree of freedom to regulate the electronic and optical properties of TBG[6-9]. Experimental and theoretical works quickly followed on the electronic structure, transport, disorder and interactions of TBG[1, 2, 10]. Tight-binding model, continuum model and density functional theory are brought up to investigate electronic band structures of TBG [8, 10, 11]. Interactions between layers lead to saddle points in band structure, which create greatly enhanced peaks in density of states (DOS) [11]. These DOS peaks are referred to as van Hove singularities [6, 10].

Tarnopolsky et al. reported a fundamental continuum model for TBG and brought up three criteria for magic angle, which are flattening of lowest bands, maximum energy gap between lowest level and lowest excited level and vanishing of Fermi velocity at Dirac point[12]. Magic angles occur in a sequence of $\theta = 1.08^\circ / n$ with $1.08^\circ$ being the first magic angle[10]. Wilson et al. developed a theory for twist angle disorder in bilayer graphene, and found the energy gaps and miniband width are strongly affected by this kind of disorder while the renormalized velocity remains unchanged[13].

Among other phenomena, TBG with magic angle hosts superconductivity, interaction-induced insulating state, quantized anomalous Hall states[2]. Cao et al. have experimentally discovered correlated insulating behavior and superconductivity at a fractional filling of an isolated narrow band on magic-angle TBG[5, 14].

They also find strange metal behavior in magic-angle TBG with near Planckian Dissipation[15]. Chen et al. investigated photonic spin Hall effect and found spin splitting in TBG experimentally. They used a theoretical framework of light-matter interaction based on surface optical conductivities [16]. Li et al. investigated thermal conductivity of TBG experimentally. They found that the thermal conductivity of TBG is lower than that of monolayer graphene and ascribed to phonons in TBG do not behave in the same way as in individual graphene layers[17]. It's no doubt that TBG enters a new phase[18].

Although electronic, optical and thermal conduction of TBG have been extensively investigated, radiation property of TBG has far less attention. Inspired by the above theoretical and experimental works, we decide to investigate thermal radiation properties of TBG. The aim is to explore the physical mechanism and find methods to realize thermal management[19, 20].

In this paper, we study optical conductivity and thermal radiation of TBG at small twist angles. We use a continuum model including more than 200 plane waves to achieve convergent energy bands. This method is valid for quite small angles. Optical conductivity of TBG with different twist angles is numerically calculated by the Kubo formula. Based on far-field radiation theory of previous works[21-23], we explore thermal radiation property of TBG. Radiation spectrum of TBG shows tunable high intensity and peak positions by changing twist angle. With magic angle, TBG radiation can be tuned to concentrate in the range of 0.05eV to 0.08eV, which is out of atmospheric transparency windows[24]. This range of electromagnetic (EM) waves are hardly transmitted in the atmosphere, such that it could not be detected by infrared (IR) cameras. Devices made of or covered with such materials are IR invisible. Such materials can also be used to fabricate textile materials for keeping warm as thermal radiation are unlikely to be transmitted through atmosphere. Our results establish magic-angle bilayer graphene as a highly tunable platform to investigate invisibility and keeping warm materials.

In section II, we introduce continuum model of TBG and investigate the magic angle from the criterion of vanishing Fermi velocity, flattening Fermi energy bands and sharp peaks in DOS. Far-field thermal radiation theory is presented in section III and numerical results of optical conductivity as well as radiative properties are investigated in section IV. TBG radiation can be tuned away from atmospheric window by changing twist angle. Finally, a conclusion is presented in section V.

## II. Magic angle in TBG

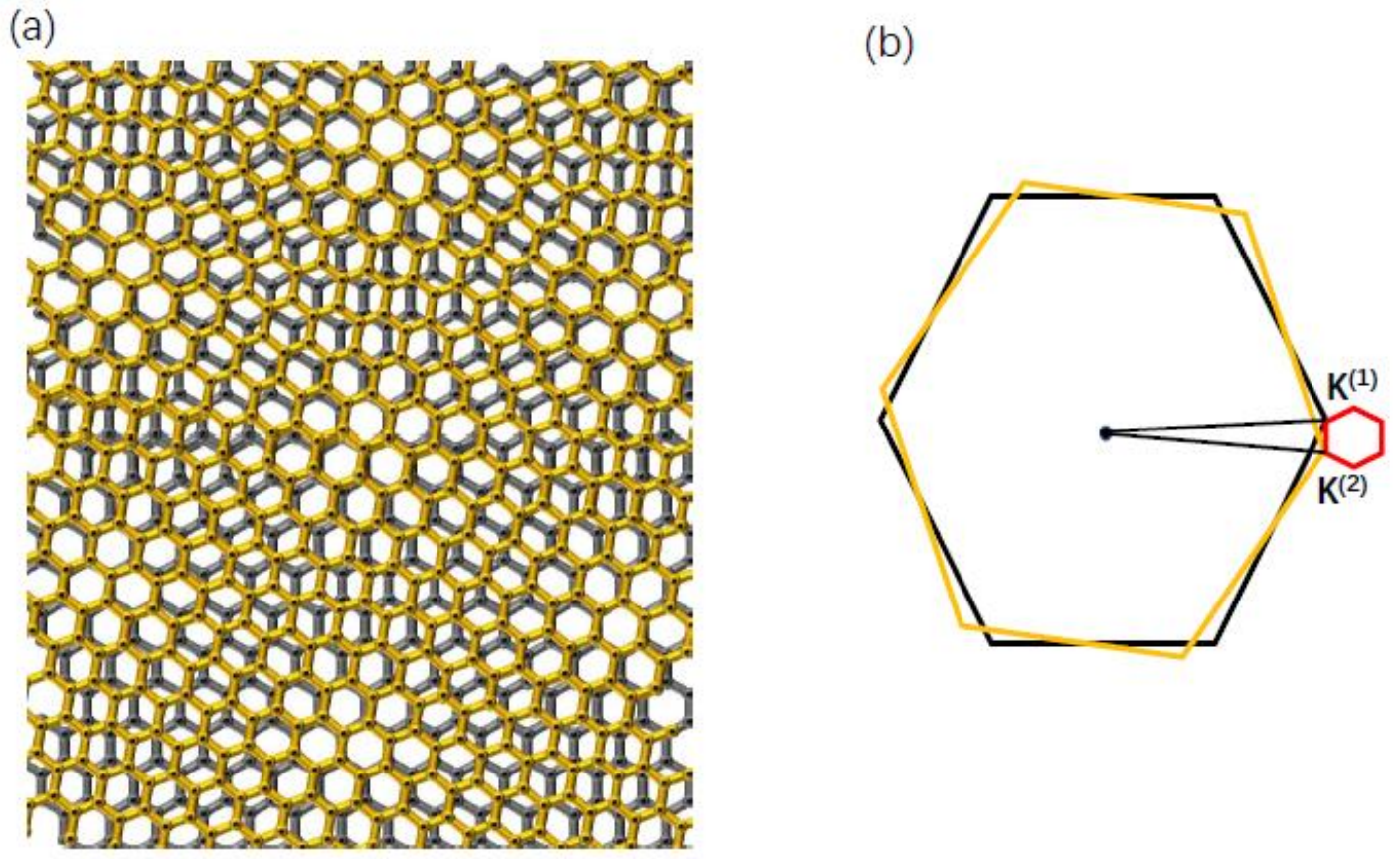

Fig. 1(a) is Moire superlattice formed by twisted bilayer graphene (TBG). (b) is the reciprocal lattice.

We start from AA-stacked bilayer graphene (i.e., perfectly overlapping honeycomb lattices) and rotate the two layers around a pair of registered A sites by $\pm\theta/2$, respectively. Moire pattern in TBG is formed. Fig.1(a) shows the TBG with a twist angle of 6.85°. In Fig.1(b), the black and yellow hexagons are Brillouin zones of the two graphene layers. They are twisted with a small angle relatively. Same Dirac points of two layers are coupled as they are very close. The red small hexagon is Brillouin zone of the Moire suplerlattice. Opposite Dirac points do not couple. Due to large Moire lattice constant, the Moire Brillouin zone is quite small.

As the Fermi velocity is angle-dependent in TBG, we first study how twist angle changes Fermi velocity. We use the continuum TBG Hamiltonian[6, 10]. It is composed of three parts: the part of layer 1, the part of layer 2 and the part of the coupling between the two layers. The low-energy effective Hamiltonian, near $K$ point, is

$$\mathcal{H}=\hbar v_F\sum_{\vec{k},\alpha,\beta}\phi^{\dagger}_{1,\vec{k},\alpha}\{\vec{\tau}^{-\theta/2}\cdot(\vec{k}+\frac{\Delta\vec{K}}{2})\}_{\alpha\beta}\phi_{1,\vec{k},\beta}+\hbar v_F\sum_{\vec{k},\alpha,\beta}\phi^{?}_{2,\vec{k},\alpha}\{\vec{\tau}^{\theta/2}\cdot(\vec{k}-\frac{\Delta\vec{K}}{2})\}_{\alpha\beta}\phi_{2,\vec{k},\beta}$$
$$+\left(\sum_{\alpha,\beta}\sum_{\vec{k},\vec{G}}\tilde{t}^{\beta\alpha}_{\perp}(\vec{G})\phi^{\dagger}_{1,\vec{k}+\vec{G},\alpha}\phi_{2,\vec{k},\beta}+H.c.\right) \quad (1)$$

Where $\vec{k}$ is the wave vector in the Moire Brillouin zone, $\hbar$ is reduced Planck constant and $v_F$ is the bare Fermi velocity of single layer graphene (SLG). The same $\vec{k}$ vector in both layers correspond to the same plane waves in the original lattices; the Dirac cones occur at $\vec{k}=\mp\Delta\vec{K}/2$ for layer 1 and layer 2 respectively, with $\Delta K=2K_D\sin\frac{\theta}{2}$ is displacement between Dirac points of the two layers while $K_D=\frac{4\pi}{3a}$ is Dirac point wave vector of SLG. $\vec{\tau}=(\tau_x,\tau_y)$, $\vec{\tau}^{\theta}=e^{+i\theta\tau_z}\vec{\tau}e^{-i\theta\tau_z}$, and $\tau_x,\tau_y$ are the Pauli matrices. $\tilde{t}^{\beta\alpha}_{\perp}(\vec{G})$ is the Fourier component of interlayer hopping, which is nonzero only at reciprocal Moire superlattice vectors of $\vec{G}$. For a particular Bloch vector $\vec{k}$ in the Moiré Brillouin zone, the interlayer coupling hybridizes the graphene's eigenstates of one layer with eigenstates in the other layer at $\vec{q}=\vec{k}+\vec{G}$. As stated in the literature [10], only three reciprocal lattice vectors $\vec{G}=0$, $\vec{G}=-\vec{G}_1$ and $\vec{G}=-\vec{G}_1-\vec{G}_2$ are selected, as the Fourier components of these vectors have relatively large amplitudes, and all other components are ignored for small amplitudes. $\phi_{i,\vec{k},\alpha}$ is the Fourier component of $i$-th layer $\alpha$ sublattice wave-function.

Explicitly, the Hamiltonian of TBG can be written as:

$$H(\vec{k})=\begin{pmatrix} \hbar v_F\vec{\tau}\cdot(\vec{k}+\frac{\Delta\vec{K}}{2}) & 0 & 0 & \tilde{t}_\perp^T(0) & \tilde{t}_\perp^T(G1) & \tilde{t}_\perp^T(G1+G2) \\ 0 & \hbar v_F\vec{\tau}\cdot(\vec{k}+\frac{\Delta\vec{K}}{2}-\vec{G}_1) & 0 & \tilde{t}_\perp^T(\vec{G}_1) & 0 & 0 \\ 0 & 0 & \hbar v_F\vec{\tau}\cdot(\vec{k}+\frac{\Delta\vec{K}}{2}-\vec{G}_1-\vec{G}_2) & \tilde{t}_\perp^T(\vec{G}_1+\vec{G}_2) & 0 & 0 \\ \tilde{t}_\perp^*(0) & \tilde{t}_\perp^*(-\vec{G}_1) & \tilde{t}_\perp^*(-\vec{G}_1-\vec{G}_2) & \hbar v_F\vec{\tau}^\theta\cdot(\vec{k}-\frac{\Delta\vec{K}}{2}) & 0 & 0 \\ \tilde{t}_\perp^*(-\vec{G}_1) & 0 & 0 & 0 & \hbar v_F\vec{\tau}^\theta\cdot(\vec{k}-\frac{\Delta\vec{K}}{2}+\vec{G}_1) & 0 \\ \tilde{t}_\perp^*(-\vec{G}_1-\vec{G}_2) & 0 & 0 & 0 & 0 & \hbar v_F\vec{\tau}^\theta\cdot(\vec{k}-\frac{\Delta\vec{K}}{2}+\vec{G}_1+\vec{G}_2) \end{pmatrix} \quad (2)$$

This is a $12\times12$ Hamiltonian which includes only 6 plane waves. However, it is not a closed system as each $\vec{k}$ point in one layer couples to three points in the other layer. In order to achieve convergence, we have to include as many plane waves as necessary. Especially in the small angle limit, when the variation of interlayer hopping becomes larger. That's a huge consumption of calculation time. In real calculation, more than hundreds of plane waves are included, and maximum size of the Hamiltonian is 450 by 450.

Electronic properties such as density of states, dispersion relations and eigenfunctions are obtained by solving the Dirac equation $H(\vec{k})\psi(\vec{k})=E(\vec{k})\psi(\vec{k})$, where $\psi(\vec{k})$ is a two-component spinor representing two inequivalent sublattices. Effective interlayer hopping 110meV and bare Fermi velocity in SLG $v_F=0.9\times10^6$ m/s are adopted in all calculation except specification otherwise[12]. The low energy levels dispersion is shown in Fig.2. Due to interlayer coupling, there are numerous energy bands for TBG. Only lowest levels are displayed in the figures. We can see condensed levels for smaller twist angle and dispersive levels for relatively large angles. Almost absolute flat bands appear when the twist angle is $\theta=1.1°$. This angle is considered as one of the magic angles for TBG.

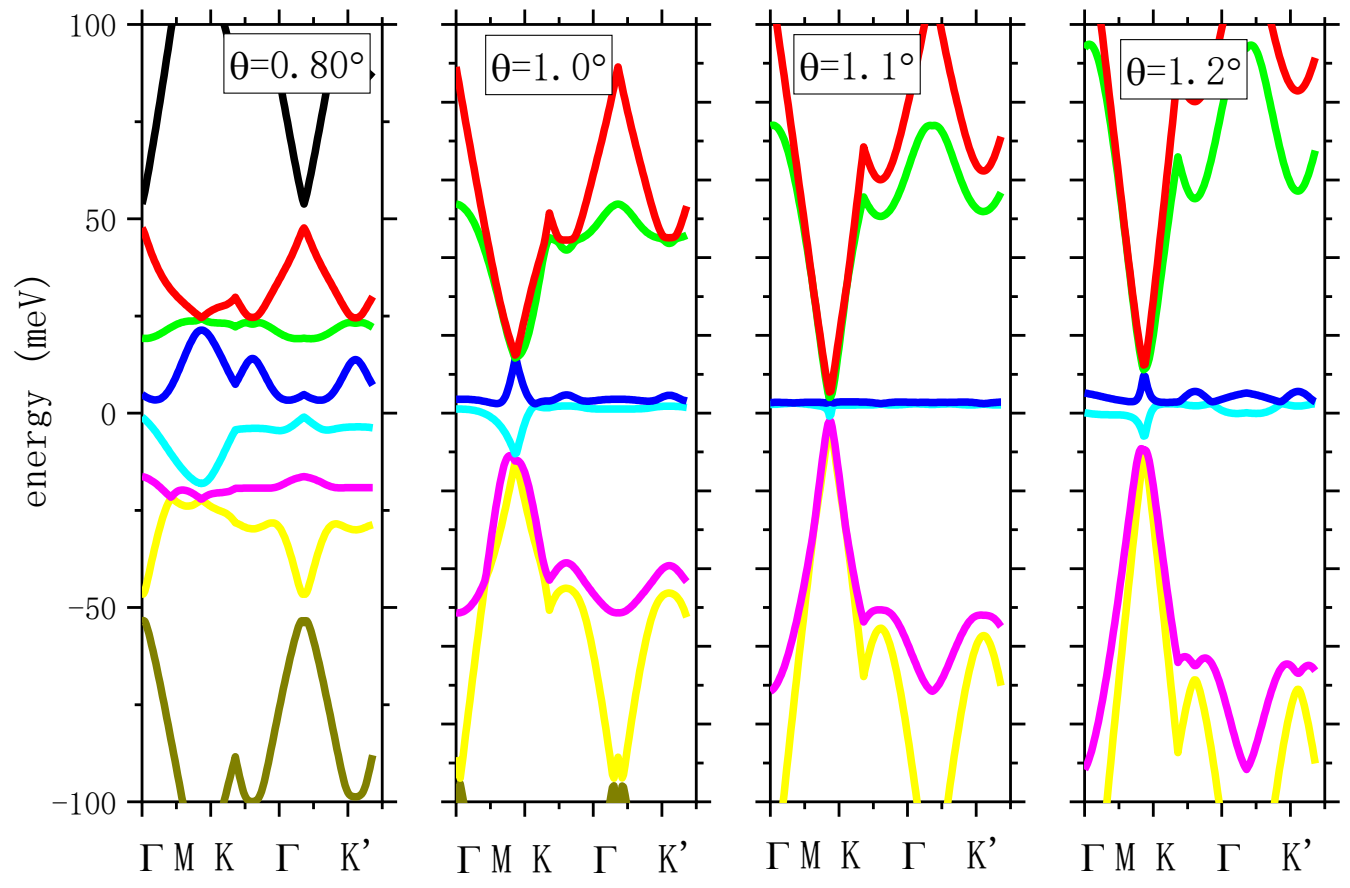

Fig.2 Dispersion of TBG for small twist angles. Only lowest levels are displayed in the plots. Effective interlayer hopping is 110meV and bare Fermi velocity in SLG is $v_F=0.9\times10^6$ m/s.

The density of states (DOS) of TBG with this magic angle $\theta=1.1°$ is plotted in Fig.3. Two giant peaks occur at the energy of $\pm t_0$ ( hopping parameter), just like the case in single layer graphene. But there appear sharp peaks in the low energy range, especially an extremely sharp peak near zero energy. These sharp peaks are induced by interlayer coupling which causes low energy levels little variation and even flat near Fermi level. This band structure of TBG is distinct from single layer graphene. Panel (b) and (c) are enlarged versions of panel (a) in order to show low energy band structures. Many van Hove singularities are observed in low energy ranges. Interlayer coupling of TBG has strong effect on low energy bands.

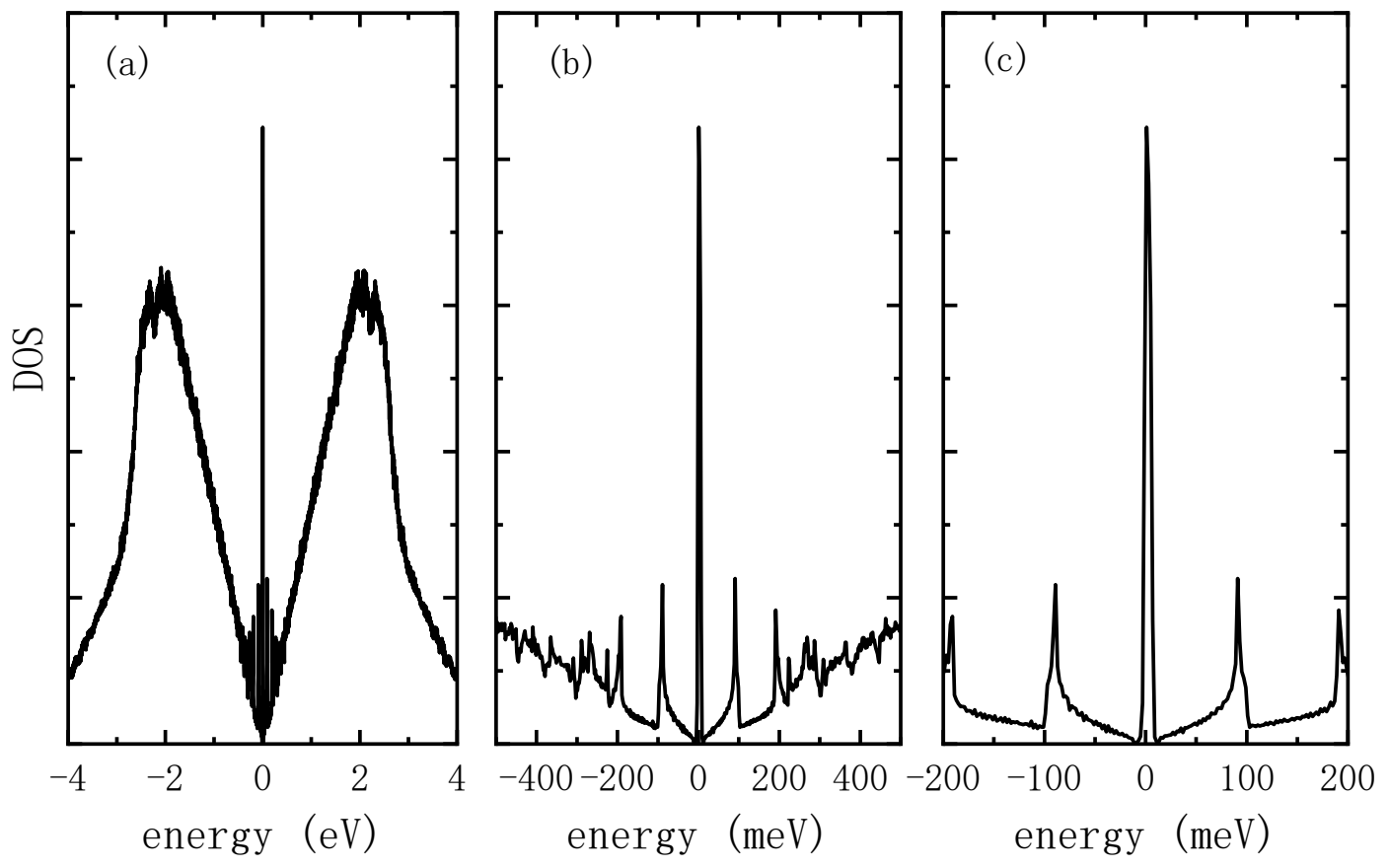

Fig.3 Density of states of TBG with twist angle $1.1°$. (a) DOS in the energy range of -4.0eV~4.0eV. (b) DOS in the energy range of -500meV~500meV. (c) DOS in the energy range of -200meV~200meV.

Then we focus on the Fermi velocity. With the eigenfunctions, the group velocity is calculated by

$$v_{n,\alpha} = \frac{1}{\hbar}\langle \psi_n | \frac{\partial H}{\partial k_\alpha} | \psi_n \rangle \tag{3}$$

Where $\psi_n$ represents eigenfunctions of the two flat bands and $\alpha = x, y$ is a Cartesian direction. So the Fermi velocity of TBG is $\tilde{v}_F = \sqrt{v_{n,x}^2 + v_{n,y}^2}$ . TBG Fermi velocity as functions of twist angle are displayed in Fig.4.

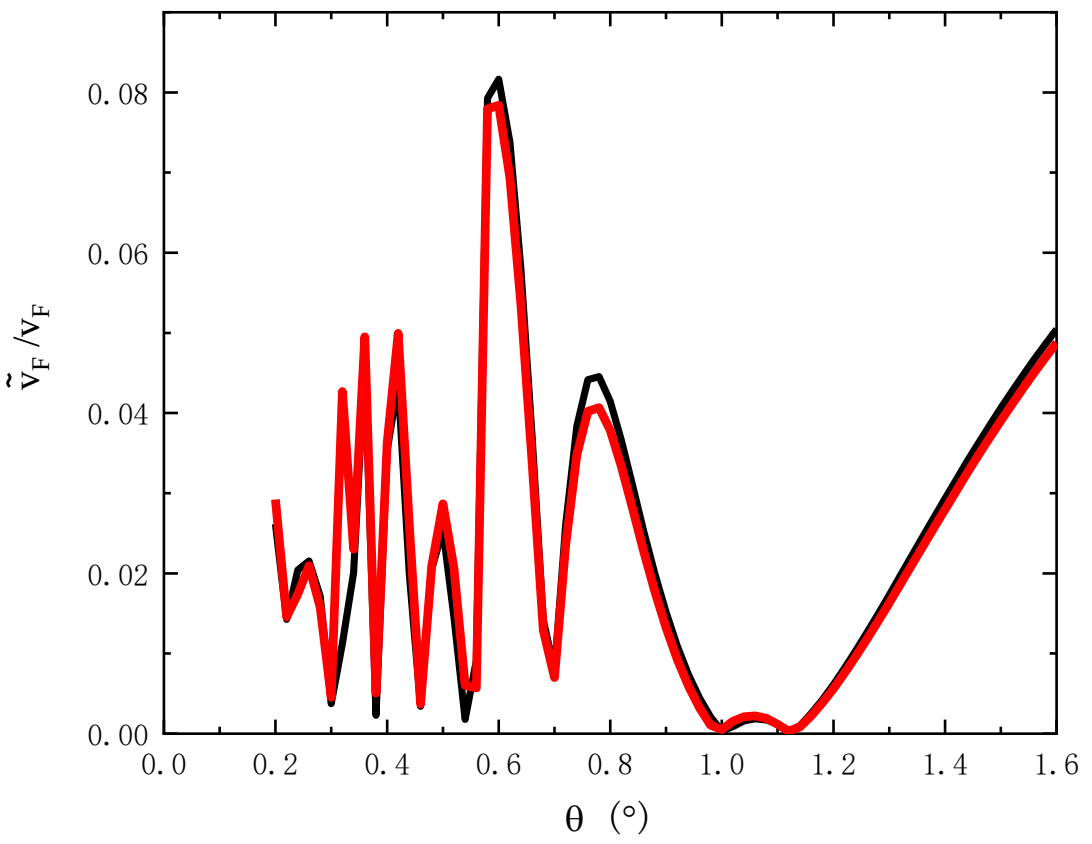

Fig.4 Fermi velocity $\tilde{v}_F$ of TBG changes with twist angle. Effective interlayer hopping is 110meV and bare

Fermi velocity in SLG is $v_F = 0.9\times10^6$ m/s. The black (red) curve represents $K$-point velocity of the flat band most close to but below (above) zero energy.

Fermi velocity of the lowest levels of TBG at Dirac point is calculated with the above formula. Due to the flat levels, Fermi velocity of TBG is quite slow at small angles. It can be seen from Fig.4 that velocity completely vanishes at angles of $1.0°$ and $1.12°$. These angles can be regarded as magic angles. There are second and third magic angles $0.54°$ and $0.38°$ with which velocity are very close to zero. All these magic angles make up a sequence which conforms to the relation of $\theta \sim \frac{1}{2\pi}\frac{t_\perp}{\hbar v_F}\frac{a}{n} \approx 1.08°\times\frac{1}{n}$ with $1.08°$ the first magic angle [10]. In fact, Fig.4 shows double values for magic angles, $1.12°$ and $1.0°$ for the 1st magic angle, $0.54°$ and $0.46°$ for the 2nd magic angle and $0.38°$ and $0.3°$ for the 3rd magic angle. They form two sequences with the relation of $\theta \sim \theta_{1st}\times\frac{1}{n}$. This explains Cao's report that the typical values of the resistivity ~ temperature slope are large near 1.08° and 1.16° respectively[15], which implies both 1.08° and 1.16° are magic angles. The first magic angles in our calculation are slightly different from Ref.[10, 15] or other literatures due to different values of $v_F$ adopted in calculation.

**III. Theory of thermal radiation of a plane geometry**

In general terms, emission refers to how the radiative energy emanates from an object, where its internal energy is converted to propagating electromagnetic waves[25, 26]. The energy flux of thermal radiation can be obtained from

$$\left\langle \vec{S} \right\rangle = \frac{1}{\mu_0}\left\langle : \vec{E}_\perp \times \vec{B} : \right\rangle \tag{4}$$

here normal order operation is aimed at removing the zero-point energy, so as to avoid divergence. Using Coulomb gauge, the electric and magnetic fields are related to vector potential by $\vec{E}_\perp = -\partial_t \vec{A},\ \vec{A} = \nabla\times\vec{A}$.

The energy radiation at far distance is $\left\langle S_\perp^i \right\rangle = \frac{1}{\mu_0}\sum_{jklm}\epsilon_{ijk}\epsilon_{klm}\frac{\partial}{\partial t}(-\frac{\partial}{\partial r_{l'}})i\hbar D_{jm}^<(\vec{r},t;\vec{r}_{l'},t')$. Here $\epsilon_{ijk}$ is the Levi Civita symbol and $i, j, l, m$ are Cartesian directions. $D_{jm}^<(\vec{r},t;\vec{r}\,',t') = \frac{1}{i\hbar}\left\langle A_m(\vec{r}\,',t')A_j(\vec{r},t)\right\rangle$ is the lesser Green's function. The retarded one is defined as $D_{\alpha\beta}^r(\vec{r},t;\vec{r}\,',t') = \frac{1}{i\hbar}\left\langle\left[A_\alpha(\vec{r},t), A_\beta(\vec{r}\,',t')\right]\right\rangle\Theta(t-t')$.

By taking the Fourier transform and incorporating contributions of all frequency photons, we have

$$\left\langle S_\perp^i \right\rangle = \frac{2}{\mu_0}\sum_{jklm}\epsilon_{ijk}\epsilon_{klm}\int_0^{+\infty}\frac{d\omega}{2\pi}\hbar\omega(-\frac{\partial}{\partial r_{l'}})D_{jm}^<(\vec{r},\vec{r}_{l'},\omega). \tag{5}$$

The material property is incorporated by the Keldysh equation $D_{\alpha\beta}^< = \sum_{\mu\nu} D_{\alpha\mu}^r \Pi_{\mu\nu}^< D_{\nu\beta}^a$. Here $D^a = [D^r]^\dagger$ is advanced Green's function and $\Pi^{<}_{\mu\nu}$ is the current-current correlation or self-energy of the system.

In real space, the correlation function is defined as $\Pi^{\alpha\beta,<}_{ll'}(t,t')=\frac{1}{i\hbar}\left\langle I_{l',\beta}(t')I_{l,\alpha}(t)\right\rangle$. Here the current in material lattice is defined $I_{l,\alpha}(t)=-\sum_{l_1l_2}c^{\dagger}_{l_1}(t)M^{l,\alpha}_{l_1l_2}c_{l_2}(t)$ with $M^{l,\alpha}_{l_1l_2}=(-\frac{ie}{2\hbar})H_{l_1l_2}(R^{\alpha}_{l_1}-R^{\alpha}_{l_2})(\delta_{l,l_1}+\delta_{l,l_2})$. Substituting the currents into self-energy, we obtain $\Pi^{\alpha\beta,<}_{ll'}(t,t')=-i\hbar\sum_{l_1l_2l_3l_4}M^{l,\alpha}_{l_1l_2}G^{<}_{l_2l_3}(t,t')M^{l',\beta}_{l_3l_4}G^{>}_{l_4l_1}(t',t)$. In which we have introduced electron Green's function $G^{>}_{l_1l_2}(t,t')=\frac{1}{i\hbar}\left\langle c_{l_1}(t)c^{\dagger}_{l_2}(t')\right\rangle$ and $G^{<}_{l_1l_2}(t,t')=-\frac{1}{i\hbar}\left\langle c^{\dagger}_{l_2}(t')c_{l_1}(t)\right\rangle$ with $c_{l_1}$ ($c^{\dagger}_{l_2}$) annihilation (creation) operator at site $l_1(l_2)$.

By performing the Fourier transform, in the reciprocal space the current takes the form

$\vec{I}_{\vec{q}}=(-\frac{e}{2})\frac{1}{\sqrt{N}}\sum_{\vec{k}}c^{\dagger}(\vec{k}+\vec{q})[\vec{V}(\vec{k}+\vec{q})+V(\vec{k})]c(\vec{k})$ with $V(\vec{k})=\frac{1}{\hbar}\nabla_{\vec{k}}H(\vec{k})$ the group velocity.

After transforming to mode space with $c(\vec{k})=\sum_n\varphi_n(\vec{k})\tilde{c}_n$, $c^{\dagger}(\vec{k})=\sum_m\tilde{c}^{\dagger}_m\varphi^{\dagger}_m(\vec{k})$, and expressing $\varphi_n(\vec{k})$ in the form of $\left|\vec{k},n\right\rangle$, the explicit form of current-current correlation in the long wave approximation is obtained

$$\Pi^{r}_{\alpha\beta}(\omega)=\frac{e^2}{A}\sum_{n,m,\vec{k}}\left\langle k,n\right|V^{\alpha}\left|k,m\right\rangle\left\langle k,m\right|V^{\beta}\left|k,n\right\rangle\frac{f_m(\vec{k})-f_n(\vec{k})}{\varepsilon_m(\vec{k})-\varepsilon_n(k)-(\hbar\omega+i\eta)}. \tag{6}$$

Here $\varepsilon_m(\vec{k})$ is the energy of $m$-th mode and $f_m(\vec{k})=1\Big/\left[\exp(\frac{\varepsilon_m(\vec{k})-\mu}{k_BT})+1\right]$ is Fermi distribution function, $\mu$ is the chemical potential, $k_B$ is the Boltzmann constant and $T$ is temperature.

The optical conductivity is related to the retarded component of the current-current correlation function as $\Pi^{r}_{\alpha\beta}(\omega)=-i\omega\sigma_{\alpha\beta}(\omega)$ [27]. Therefore, optical conductivity can be expressed as

$$\sigma_{\alpha\beta}(\omega)=\frac{i}{\omega}\frac{e^2}{N\Omega}\sum_{mn,k}\frac{\left\langle\vec{k},n\right|V^{\alpha}(\vec{k})\left|\vec{k},m\right\rangle\left\langle\vec{k},m\right|V^{\beta}(\vec{k})\left|\vec{k},n\right\rangle}{\varepsilon_m(\vec{k})-\varepsilon_n(\vec{k})-(\hbar\omega+i\eta)}\left[f_m(\vec{k})-f_n(\vec{k})\right]. \tag{7}$$

By inserting these formulae into energy flux in Eq. (5) we obtain total energy radiation. For the plane geometry of material, in the far place, $\left\langle S^x\right\rangle=\left\langle S^y\right\rangle=0$. So finally, the radiation is [22] $\left\langle S^z\right\rangle=\int_0^{\omega}S_z(\omega)d\omega$,

where

$$\langle S^z(\omega)\rangle = \frac{2}{3\pi\varepsilon_0 c^3}\hbar\omega^3 N(\omega)\mathrm{Re}(\sigma_{xx}(\omega)) \tag{8}$$

In which $N(\omega)=1/[\exp(\hbar\omega/k_B T)-1]$ is Bose distribution function.

## IV. Numerical results

Optical conductivity represents property of material and plays import role in radiation. Firstly, we calculate the conductivity as well as radiation spectrum of TBG. Fig.5 is the result at temperature of $300$ K.

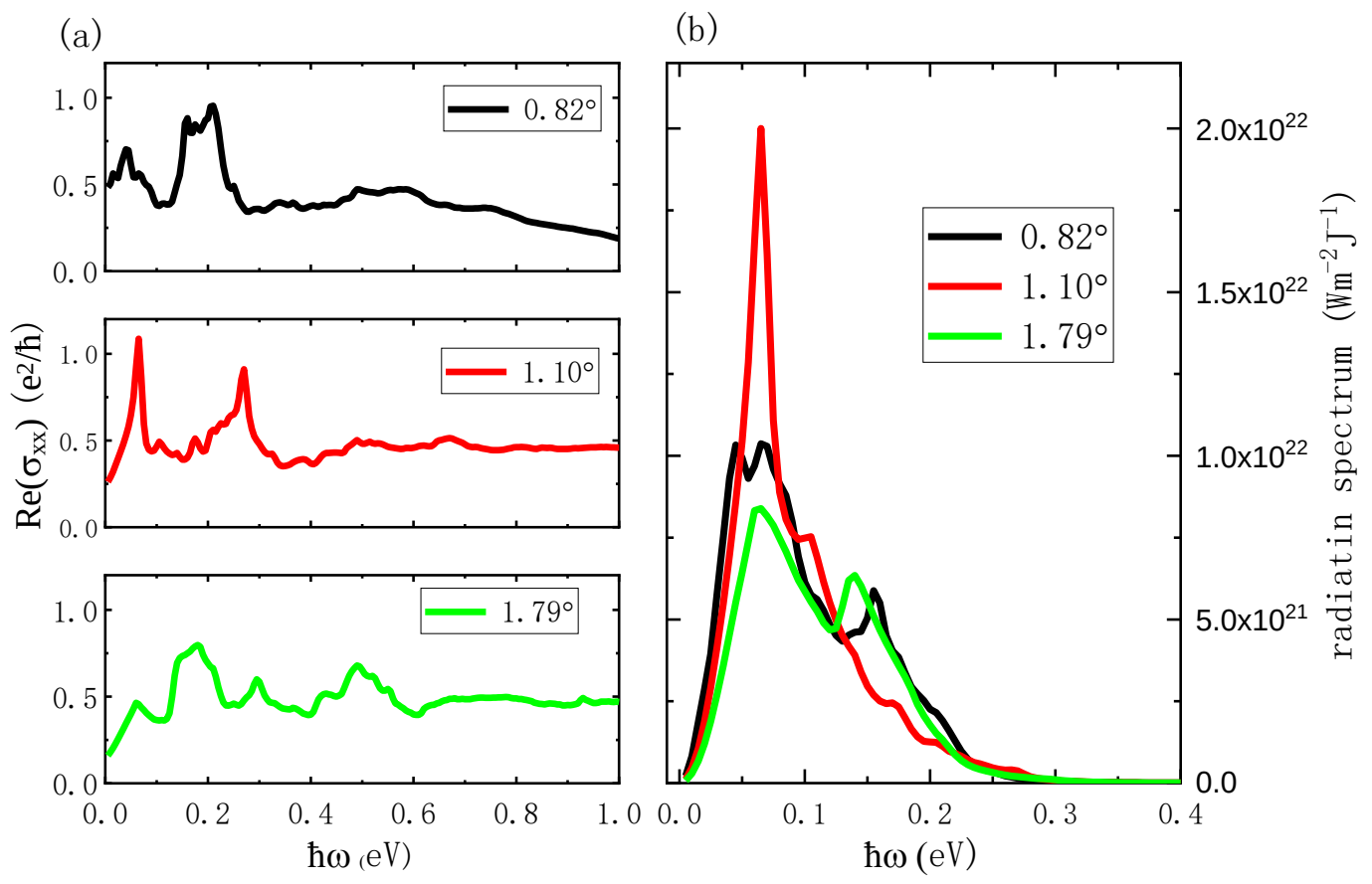

Fig.5 (a) Optical conductivity and (b) Radiation power of TBG with different twist angles change with frequency of photon energy (T=300K).

Generally, optical conductivities change with $\omega$ or photon energy with peaks at low frequency and approach constant at high frequency. The value of flat part is roughly $\sigma_{xx}=0.5e^2/\hbar$, which is consistent with Ref[28]. This value is twice of the single layer graphene conductivity with $\sigma_0=0.25e^2/\hbar$ [29]. This implies that at low frequency twist angle has great effect on TBG properties, while at high frequency TBG exhibits like two isolated graphene layers. The prominent peaks appear at low frequency of TBG with $\theta=1.1°$. They reflect electron transitions between van Hove singularities in the plot of density of states, as shown in Fig.3 . The low frequency peaks for TBG with twist angle of $0.82°$ and $1.79°$ are a little squatty. This is due to strong interlayer coupling for small angles. Interlayer coupling varies rigously with twist angle, especially for small angles.

Conductivity for twisted angle of $0.82°$ drops a little when frequency is more than half of 1.0eV. Because for small angles, more levels concentrate in low energy range. In calculation we cannot take all those bands into consideration for limited calculation power. This doesn't affect our study of radiation, as high frequency conductivity contribute infinitesimally small, as can be seen in the radiation spectrum plot. Fig.5(b) is radiation spectrum of TBG with different twisted angles. The radiation spectrum of TBG is not as smooth as that of black body. There are main peak and side peaks with any twist angle. The main peak of TBG with magic angle $\theta=1.1°$ is sharper than other cases, corresponding to electron transitions between zero energy flat band and low energy bands. The full width at half maximum (FWHM) of radiation with $\theta=1.1°$ is between $0.05 \sim 0.08eV$ , which converted to wavelength between $15 \sim 24\mu m$ . This wavelength range is out of atmospheric window. This means radiative waves from 1.1° TBG at room temperature are unlikely transmitted in atmosphere. In other words, it is hardly detected by IR camera from a certain distance so as to realize aim of invisibility. Basing on this point, TBG could be one of the ideal stealth materials to make some devices infrared invisible. It can also be used to make fabric for keeping warm. On the other hand, as such waves are unlikely dissipated by atmosphere, electronic devices made of TBG are difficult to cool down. This difficult situation could be solved by changing twist angle to 1.79° or larger, so the radiation is moved to atmospheric window.

Since twist angle $\theta = 1.1°$ is supposed to be magic angle of TBG, we explore more of TBG with this angle. Conductivity of TBG at different temperatures is displayed inFig.6(a).

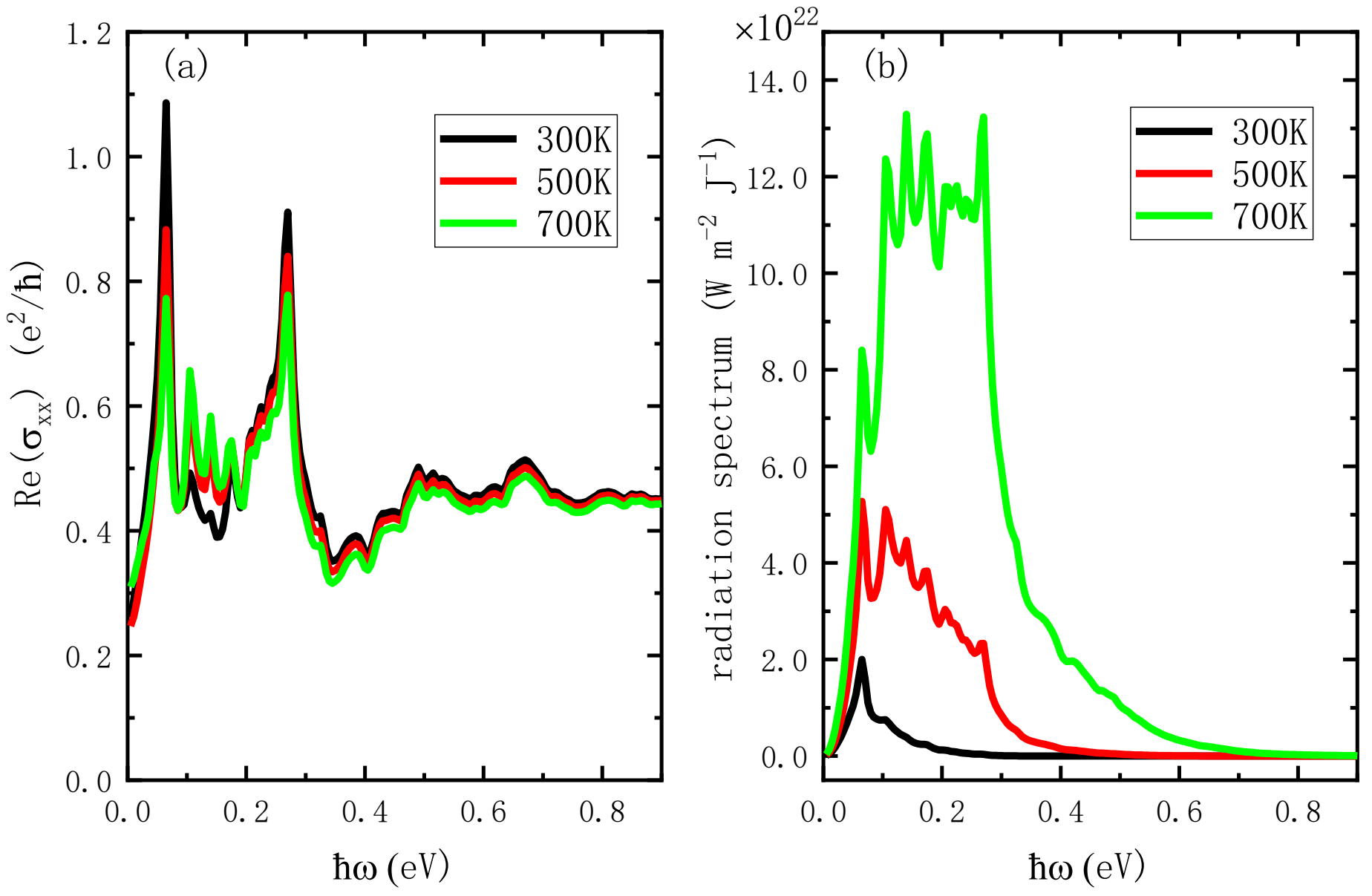

Fig.6 (a)Optical conductivity and (b)radiation spectrum of TBG with twist angle of $\theta = 1.1°$ at different temperatures.

Main features happen at low frequency range. Peak positions are almost the same at different temperatures. When temperature increases, peak values drops and small peaks appear between the two giant peaks. Fig.6(b) is radiation spectrum of magic angle TBG. Radiation increases greatly with temperature. The maximum values move to higher frequency, roughly reflecting Wien's displacement law. There are small structures in each of the curves, which deviate from Planckian black body radiation. However, these deviations are averaged when integrating over frequency. Therefore, total radiation of TBG is close to the black body radiation with straight lines in the log-log plot, see Fig.7.

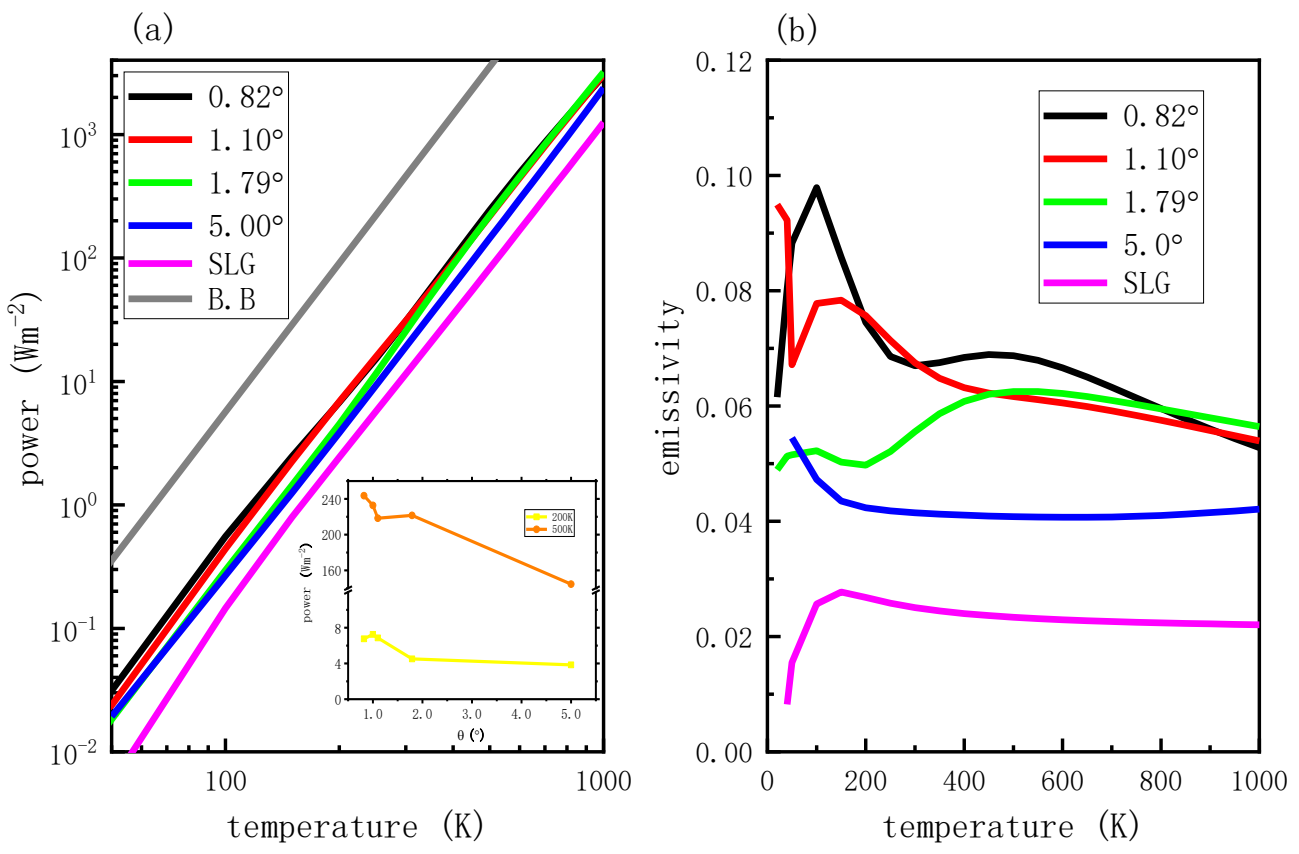

Fig.7 (a)Total radiation of small angle TBG vs temperature, by comparison with black body and single layer graphene. (b) Emissivity of TBG with different twist angles.

Fig.7(a) is the total radiation power of TBG with four different twist angles change with temperature. In this log-log plot, radiation of TBG with small twist angles are roughly straight lines as temperature increases. These lines are close to each other, implying small differences in radiation with twist angles. The grey line represents black body radiation. TBG radiation lines parallel to that of black body, signifying radiation of small angles TBG obey $T^4$ law. The purple line is radiation of single layer graphene, which also observes $T^4$ law except at low temperature range. The inset of Fig.7(a) reveals how radiation changes with twist angle. No matter at 200K or 500K, radiation generally decreases with twist angle, with small fluctuation for quite small angles. This reflects strong coupling between layers for small twist angles, while for bigger twist angles, interlayer coupling nearly disappears. As stated above, when $T = 200\,\text{K}$, bilayer graphene with twist angle of $1.0°$ produces highest power. At temperature of $T = 500\,\text{K}$, TBG with twist angle of $1.1°$ radiates relatively less compared with $1.0°$ or $1.2°$.

Both TBG and SLG radiation are far less than a black body. If we define emissivity as the ratio of radiation of a practical object to the ideal black body [30] , we can plot emissivity of TBG accordingly. This would also clearly show radiation property of bilayer graphene of different angles, as displayed in Fig.7(b). They are distinct at low temperatures, but approach identical as temperature increases, reflecting interlayer coupling sensitive to small twist angle. The emissivity of TBG and SLG change vastly when temperature is lower than 300K. But at high temperature range, the emissivity of TBG is roughly twice of that of single layer graphene. The value of TBG emissivity is around 0.05, signifying TBG kind of metal.

Radiation of TBG changes when doped with electrons or holes. Fig.8 is radiation of TBG as function of twist angle with different chemical potentials at 300K.

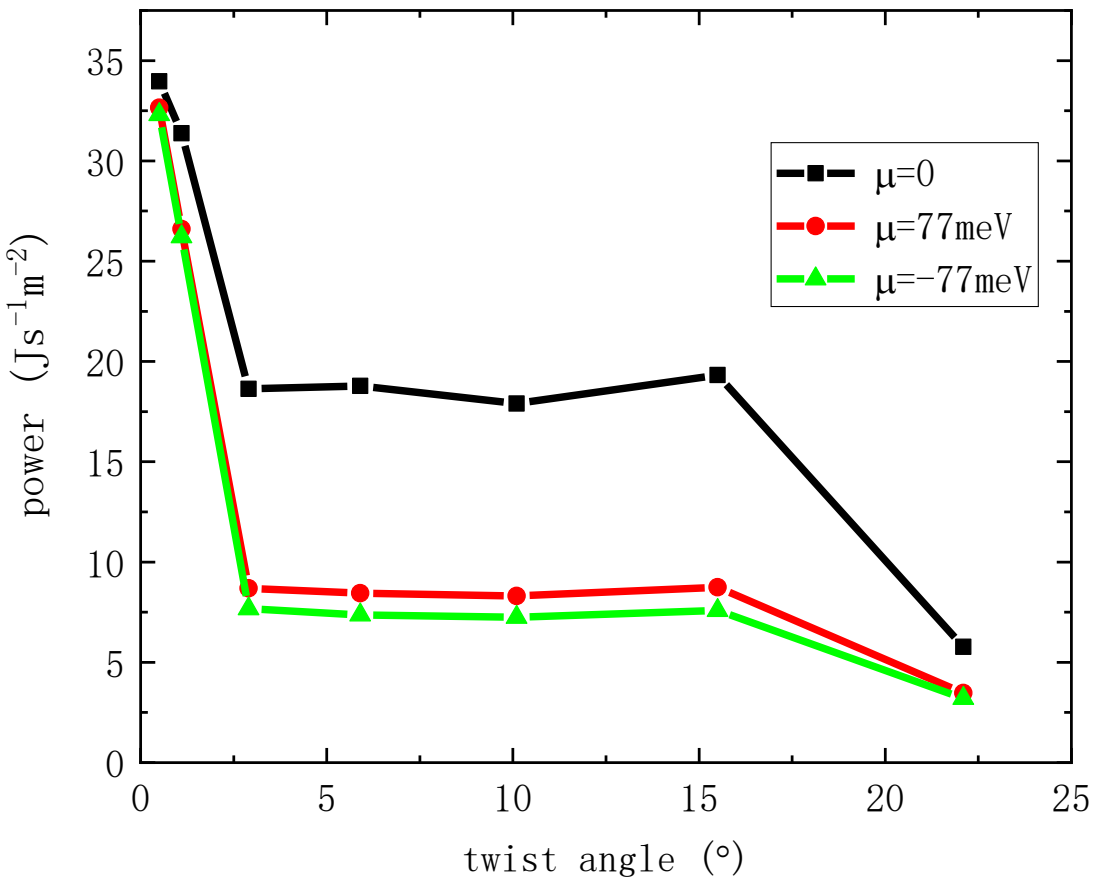

Fig.8 Total radiation of TBG as a function of twist angle（T=300K）.

The black curve is for neutrality TBG. Radiation decreases with twist angle when the angle is quite small or relatively large. But for medium angles, radiation is almost constant. This implies the interlayer coupling does not vary in this range. For small angles, interlayer coupling is strong. If the TBG is doped, no matter with electrons or holes, radiation decreased a lot. As doping results in chemical potential raised or lowed, electrons in some of the lowest levels are prohibited from transitions.

In the above investigation, effective interlayer coupling $w_1$ is taken to be 110meV [11], which is widely used in tight-binding model or continuum models [6, 10, 12]. However, the effective coupling varies with the interlayer distance. Conventionally the distance between layers is 0.335nm, as in graphite. When the distance in TBG is increased, the coupling becomes weak. Fig.9 investigates how the coupling affect radiation of TBG with twist angle of $1.0°$. Radiation increases with coupling, no matter at high temperatures or low temperatures. $T^4$ law is always satisfied. The inset is log-log plot of TBG radiation with different coupling, showing more details at low temperatures. It can be seen clearly that TBG emits more heat energy when coupling is strong, even at low temperatures. This suggests controlling measures include decreasing twist angle and increasing interlay coupling in order to achieve more radiation.

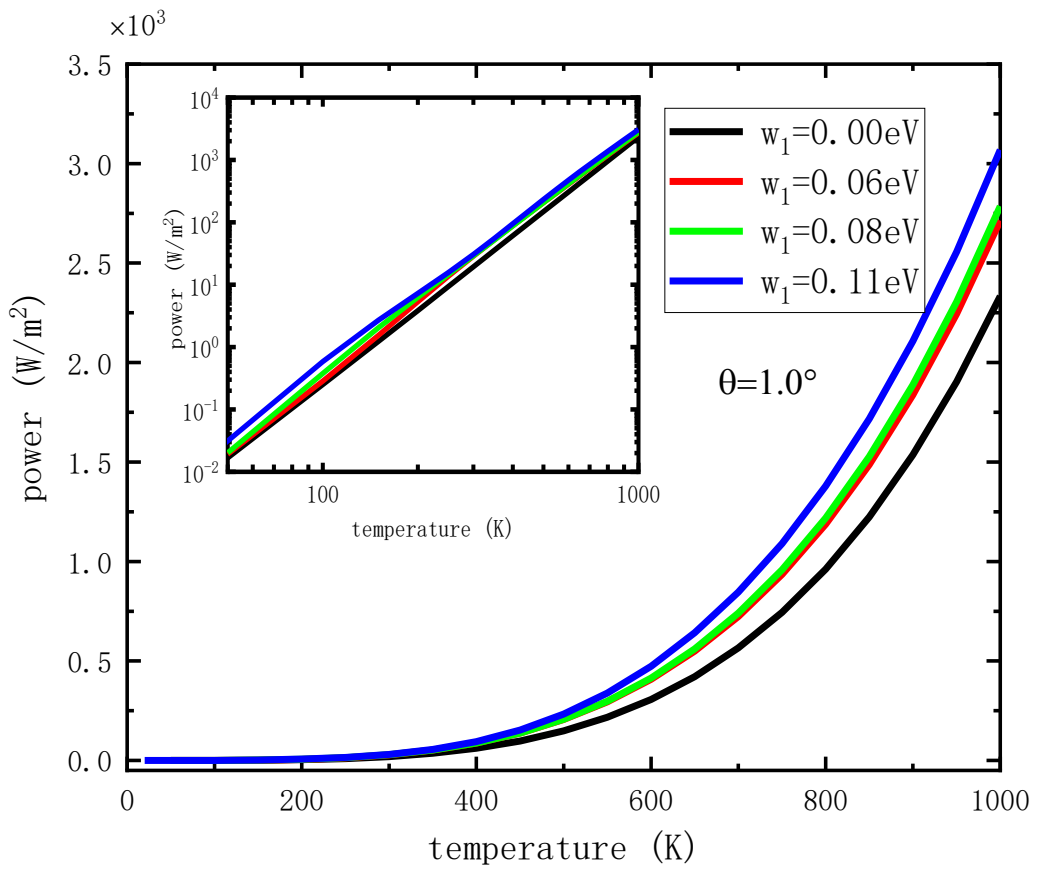

Fig.9 Total radiation of magic angle TBG with different interlayer coupling. The twist angle is 1.0°.

**V. Conclusion:**

We theoretically investigate Fermi velocity of TBG to determine magic angles. The presence of interlayer interactions in twisted bilayer graphene (TBG) enhances characteristic of the optical and electronic properties. The thermal radiation spectrum can be tuned away from atmospheric window by changing twist angles. The peculiar radiation spectrum can be of great use in invisibility and keeping warm. The radiation spectrum of TBG is not smooth, which is different from black body and grey body radiation. The captivating property originates from interlayer interactions and can also be obtained by doping electrons or holes to TBG or varying distance between layers to alter the effective coupling strength.

Acknowledgement

We acknowledge the support of MOE tier 2 grant R-144-000-411-112.

[1] Nimbalkar, A. and H. Kim, *Opportunities and Challenges in Twisted Bilayer Graphene: A Review.* Nanomicro Lett, 2020. **12**(1): p. 126.

[2] Andrei, E.Y. and A.H. MacDonald, *Graphene bilayers with a twist.* Nat Mater, 2020. **19**(12): p. 1265-1275.

[3] Koshino, M., et al., *Maximally Localized Wannier Orbitals and the Extended Hubbard Model for Twisted Bilayer Graphene.* Physical Review X, 2018. **8**(3): p. 031087.

[4] Lian, B., Z. Wang, and B.A. Bernevig, *Twisted Bilayer Graphene: A Phonon-Driven Superconductor.* Phys Rev Lett, 2019. **122**(25): p. 257002.

[5] Cao, Y., et al., *Unconventional superconductivity in magic-angle graphene superlattices.* Nature, 2018. **556**(7699): p. 43-50.

[6] Bistritzer, R. and A.H. MacDonald, *Moire bands in twisted double-layer graphene.* Proc Natl Acad Sci U S A, 2011. **108**(30): p. 12233-7.

[7] Cao, Y., et al., *Tunable correlated states and spin-polarized phases in twisted bilayer-bilayer graphene.* Nature, 2020. **583**(7815): p. 215-220.

[8] Lopes Dos Santos, J.M., N.M. Peres, and A.H. Castro Neto, *Graphene bilayer with a twist: electronic structure.* Phys Rev Lett, 2007. **99**(25): p. 256802.

[9] Xu, X., et al., *Single-valley state in a two-dimensional antiferromagnetic lattice.* Physical Review B, 2021. **104**(20): p. 205430.

[10] Lopes dos Santos, J.M.B., N.M.R. Peres, and A.H. Castro Neto, *Continuum model of the twisted graphene bilayer*, in *Physical Review B*. 2012. p. 155449.

[11] Carr, S., et al., *Exact continuum model for low-energy electronic states of twisted bilayer graphene.* Physical Review Research, 2019. **1**(1): p. 013001.

[12] Tarnopolsky, G., A.J. Kruchkov, and A. Vishwanath, *Origin of Magic Angles in Twisted Bilayer Graphene.* Phys Rev Lett, 2019. **122**(10): p. 106405.

[13] Wilson, J.H., et al., *Disorder in twisted bilayer graphene.* Physical Review Research, 2020. **2**(2): p. 023325.

[14] Cao, Y., et al., *Correlated insulator behaviour at half-filling in magic-angle graphene superlattices.* Nature, 2018. **556**(7699): p. 80-84.

[15] Cao, Y., et al., *Strange Metal in Magic-Angle Graphene with near Planckian Dissipation.* Phys Rev Lett, 2020. **124**(7): p. 076801.

[16] Chen, Z., et al., *Photonic spin Hall effect in twisted bilayer graphene.* J Opt Soc Am A Opt Image Sci Vis, 2021. **38**(8): p. 1232-1236.

[17] Li, H., et al., *Thermal conductivity of twisted bilayer graphene.* Nanoscale, 2014. **6**(22): p. 13402-8.

[18] Hill, H.M., *Twisted bilayer graphene enters a new phase.* Physics Today, 2020. **73**(1): p. 18-20.

[19] He, M., et al., *Active control of near-field radiative heat transfer by a graphene-gratings coating-twisting method.* Opt Lett, 2020. **45**(10): p. 2914-2917.

[20] He, M.-J., et al., *Magnetoplasmonic manipulation of nanoscale thermal radiation using twisted graphene gratings.* International Journal of Heat and Mass Transfer, 2020. **150**: p. 119305.

[21] Zhang, Z.-Q., J.-T. Lü, and J.-S. Wang, *Angular momentum radiation from current-carrying molecular junctions.* Physical Review B, 2020. **101**(16): p. 161406(R).

[22] Zhang, Y.M. and J.S. Wang, *Far-field heat and angular momentum radiation of the Haldane model.* J Phys Condens Matter, 2020. **33**: p. 055301.

[23] Zhu, T., M. Antezza, and J.-S. Wang, *Dynamical polarizability of graphene with spatial dispersion.* Physical Review B, 2021. **103**(12): p. 125421.

[24] Zhai, Y., et al., *Scalable-manufactured randomized glass-polymer hybrid metamaterial for daytime radiative cooling.* Science, 2017. **355** p. 1062–1066

[25] Francoeur, M. and M. Pinar Mengüç, *Role of fluctuational electrodynamics in near-field radiative heat transfer.* Journal of Quantitative Spectroscopy and Radiative Transfer, 2008. **109**(2): p. 280-293.

[26] Zhou, C.-L., et al., *Polariton topological transition effects on radiative heat transfer.* Physical Review B, 2021. **103**(15): p. 155404.

[27] Zhang, Z.-Q. and J.-S. Wang, *Electroluminescence and thermal radiation from metallic armchair carbon nanotubes with defects.* Physical Review B, 2021. **104**(8): p. 085422.

[28] Stauber, T., P. San-Jose, and L. Brey, *Optical conductivity, Drude weight and plasmons in twisted graphene bilayers.* New Journal of Physics, 2013. **15**(11): p. 113050.

[29] Falkovsky, L.A., *Optical properties of graphene.* Journal of Physics: Conference Series, 2008. **129**: p. 012004.

[30] Freitag, M., et al., *Thermal infrared emission in graphene.* Nature Nanotechnology, 2010. **5**(7): p. 497.